# A New Transmitted Reference Pulse Cluster Based Ultra-Wideband Transmitter Design

Yiming Huo[a], Xiaodai Dong[a], Ping Lu[b]

[a] DepaElectrical and Computer Engineering Department, University of Victoria, Canada
[b] Department of Information Technology, LTH, Lund University, Sweden

**ABSTRACT**

An energy efficient ultra-wideband (UWB) transmitter based on the novel transmitted reference pulse cluster (TRPC) modulation scheme is presented. The TRPC-UWB transmitter integrates, namely, wideband active baluns, wideband I-Q modulator based up-conversion mixer, and differential to single-ended converter. The integrated circuits of TRPC-UWB front end is designed and implemented in the 130-nm CMOS process technology. the measured worst-case carrier leakage suppression is 22.4 dBc, while the single sideband suppression is higher than 31.6 dBc, operating at the frequency from 3.1 GHz to 8.2 GHz. With adjustable data rate of 10 to 300 Mbps, the transmitter achieves a high energy efficiency of 38.4 pJ/pulse.

*Keywords*: Ultra-wideband (UWB); Transmitted reference pulse cluster; I-Q modulator; CMOS

## 1. Introduction

Spectrum is an increasingly scarce resource as the wireless communicaiton market prospers with an unprecedented speed. The spectrum below 6 GHz becomes very crowded. UWB technology has emerged as a candidate solution to short-range high data rate communications thanks to its ultra-wide spectrum in the unlicensed 3.1-10.6 GHz band allocated by the Federal Communicatons Commission (FCC). Impulse radio (IR) is widely used for UWB communications. In the literature, IR-UWB transmitters can be realized by the structrues in [1], [2] or without a carrier [3]. In these IR-UWB designs, simple modulation schemes such as on-off keying (OOK), binary phase-shift keying (BPSK), and pulse-position modulation (PPM), are employed, with very good energy efficiency demo. IR-UWB can be categorized into coherent and non-coherent schemes, and non-coherent schemes generally have much lower complexity, power consumption and cost than coherent without the need to estimate the long multipath UWB channels at moderate price of error performance or data rate. Among the non-coherent (NC) schemes, transmitted reference pulse cluster (TRPC) leads in error rate performance, data rate, robustness and ease of implementation [4]. Furthermore, it provides robust performance to time variation and immunity to pulse distortion which is caused by frequency dependent antenna and channel effects. The baseband equivalent TRPC transmitted signal is

$$\tilde{s}(t) = \sqrt{\frac{E_b}{2N_f}} \sum_{m=-\infty}^{\infty} \sum_{i=0}^{N_f-1} [g(t-mT_s-2iT_d) + b_m g(t-mT_s-2iT_d)] \quad (1)$$

where $N_p$ stands for the number of total pulses in one cluster, $E_b$ indicates the average energy per bit, $g(t)$ is the component pulse of width $T_p$ in a pulse cluster, $T_s$ is the symbol duration, and $T_d$ is the delay among the pulses in one cluster. Usually $T_d = T_p$ or a few multiples of $T_p$, and $T_s \geq N_p T_d + \tau_{max}$, where $\tau_{max}$ denotes the maximum channel delay. Fig. 1 illustrates TRPC signal structures, symbol "1" is represented by one pulse cluster which consists of all "positive" pulses, and the symbol "0" is represented by one pulse cluster containing both positive and negative pulses. A simple auto-correlation receiver with short delay lines and low analog-to-digital sampling rate can successfully collect the transmitted energy that is spread into multipaths, without explicit channel estimation. It has been shown in [4] that TRPC achieves 2-3 dB and 1.3-2 dB power gain over conventional TR and NC-PPM schemes. In this paper, we present a novel TRPC-UWB transmitter design based on the CMOS transmitter front-end in a conference abstract [5]. Section 2 describes the transceiver system architecture and derives the transmitter specifications to comply with the FCC UWB emission limit. In Section 3 detailed circuit deisgn is presented. Section 4 presents experimental verification results that demonstrate good RF and energy efficiency performance. Conclusions are drawn in Section 5.

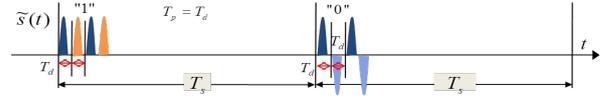

Fig. 1. TRPC structure.

## 2. TRPC-UWB transmitter specification

Fig. 2 illustrates the main function blocks of a TRPC-UWB transceiver. After the pulse shaping, the identical TRPC pulse train $s(t)$ is sent to both I and Q branches at the tranmistter end. At the receiver end, the UWB signal is down-converted through an I-Q demodulator followed by low-pass filters (LPFs) and auto-correlators. After signal combing, pulse integration and decimation, the baseband signal is successfully recovered. The mathematical derivation proves that the constant carrier frequency successfully cancelled by using TRPC scheme and I-Q modulation/demodulation architecture [6]. Therefore TRPC-UWB system is not sensitive to the frequency and phase mismatch. Another benefit of using the direct-conversion topology is that it has highly pure output without undesired frequency products [7]. These advantages obtained from system-level largely alleviate the complexity and difficulty of implementation.

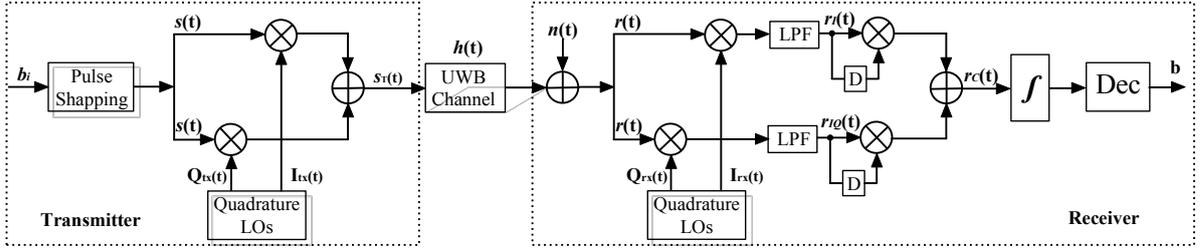

Fig. 2. Block diagram of the proposed TRPC-UWB transceiver.

Furthermore, FCC regulates that, the effective isotropic radiated power (EIRP) should not excede -41.25 dBm/MHz, and the peak EIRP density level should be below 0 dBm in a 50 MHz bandwidth, so that the unlicensed UWB signals will not interfere with other spectrum. For the TRPC signaling structure, it remains to find out the component pulse amplitude that leads to EIRP meeting FCC regulation. In a UWB system which employs a single pulse with pulse width $T_p$ and pulse repetition frequency (PRF) $R_p$, the relationship between the entire power of full bandwidth (FBW) peak power and the average power of the UWB signal is indicated by the equation as the following

$$P_{ave} = P_{peak} \cdot \delta \tag{2}$$

where $P_{peak}$ is the FBW peak power, $\delta = T_p R_p$ is the pulse duty cycle. However, due to limited resolution bandwidth (RBW) in the spectrum analyzer measurements, the measured peak and average power vary from the theoretical calculations above. According to [8], a RBW filter with bandwidth BR leads to

$$P_{ave}^m = P_{peak}^m = (R_P \cdot \tau_R)^2 \cdot P_{peak} \cdot T_P^2 \cdot B_R^2 = P_{peak} \cdot T_P^2 \cdot R_P^2$$
$$R_P \gg B_R \tag{3}$$

where $P_{ave}^m$ and $P_{peak}^m$ are the measured average and peak power, and $\tau_R$ is the reciprocal of $B_R$. Eq. (3) implies that when $R_p \gg B_R$, the RBW filter in the spectrum analyzer effectively sums $R_p \cdot \tau_R$ pulses, and consequently the amplitude increases by $R_p \cdot \tau_R$ times, and the power $(R_p \tau_R)^2$ times. On the other hand, TRPC signaling has a unique structure consisting of $N_p$ contiguous pulses and the cluster repeats with symbol rate $R$ which is much larger than $B_R$. Following the similar argument in [12], the output of the spectrum analyzer is the sum of the $N_p \cdot R \cdot \tau_R$ component pulses. Hence the measured average and peak power are given by

$$P_{ave}^m = P_{peak}^m = (N_P \cdot R \cdot \tau_R)^2 \cdot P_{peak} \cdot T_P^2 \cdot B_R^2 = N_P^2 \cdot P_{peak} \cdot T_P^2 \cdot R^2$$
$$R \gg B_R. \tag{4}$$

In this work, the designed data rate ranges from 10 Mbps to 300 Mbps, which is much larger than RBW of 1 MHz. Moreover, (4) indicates the measured power will increase by 6.02 dB when the number of the pulses is doubled. By referring to FCC UWB emission limit, the measured average EIRP and peak EIRP should meet the requirement below:

$$P_{ave}^m \leq -41.25 dBm \text{ or } 75 \text{ nW in 1 MHz RBW} \tag{5}$$

$$P_{peak}^m \leq \left(\frac{B_R}{50 \times 10^6}\right) \text{mW or } 10^6 \leq B_R \leq 50 \times 10^6. \tag{6}$$

It is apparent that the TRPC-UWB transmission power is FCC average power constrained and we can calculate $P_{peak}$ to ensure (4) meets the condition in (5). Next the amplitude of a component pulse in TRPC is found from the obtained $P_{peak}$ and the actual pulse shape. In this work, the TRPC signaling employs root raised cosine (RRC) pulses to form a cluster. The normalized RRC pulse is given by

$$g(t) = \frac{\cos[(1+\beta)\pi t/T] + \dfrac{\sin[(1-\beta)\pi t/T]}{4\beta t/T}}{1-(4\beta t/T)^2} \tag{7}$$

where $\beta$ is roll-off factor and equals to 0.25 in this work. When $g(t)$ is modulated by a LO signal, the up-converted signal has the time-domain expression as

$$s(t) = A_{TX} \cdot g(t) \cdot \cos(\omega_{LO} \cdot t) \tag{8}$$

where $A_{TX}$ is the amplitude of the output carrier modulated pulse, which depends on the overall gain of the entire transmitter, the pulse amplitude and the strength of the LOs signal. The FBW power peak of this transmitted TRPC component pulse is formulated as

$$P_{Peak,RRC} = \int_{-T_P/2}^{T_P/2} \frac{s^2(t)}{Z_{load} \cdot T_p} dt \tag{9}$$

where $Z_{load}$ is the load impedance of instrument or antenna. Finally, the exact amplitude $A_{TX}$ of TRPC-UWB output signal that fully complies with the FCC UWB emission limit is derived and summarized in Table I. The maximum amplitudes of the output signal indicate that the overall transmitter gain should be comparatively small and tunable, considering that the typical peak to peak amplitudes of LOs signals are 1 V.

## 3. Detailed transmitter design

As shown in Fig. 3, the baseband pulse cluster which occupies a bandwidth from DC to more than 1200 MHz are directly fed into the wideband active baluns respectively on I and Q paths. The double-balanced up-conversion mixers based I-Q modulator. Finally, the up converted differential RF signals are transformed to single-ended ones through a differential to single-ended (D-to-S) converter with a variable gain control. The dc offset in the baseband input leads to the growth of carrier leakage, which does not only deteriorate the error vector magnitude (EVM) but also elevates the emission level close to the FCC UWB mask.

Table 1 Pulse cluster design specification

| Data Rate (Mbps) | $N_P$ | $T_P$ (ns) | $BW_{3-dB}$ (MHz) | $P_{Peak, RRC}$ (dBm) | Max. Amp. (mV) |
|---|---|---|---|---|---|
| 10 | 8 | 1.65 | 650 | -23.47 | 32.8 |
| 20 | 8 | 1.65 | 650 | -29.47 | 16.4 |
| 40 | 8 | 0.85 | 1180 | -29.47 | 16.4 |
| 100 | 4 | 0.85 | 1180 | -31.93 | 12.36 |
| 200 | 4 | 0.85 | 1180 | -37.9 | 6.21 |
| 250 | 3 | 0.85 | 1180 | -37.3 | 6.63 |
| 300 | 2 | 0.85 | 1180 | -35.4 | 8.28 |

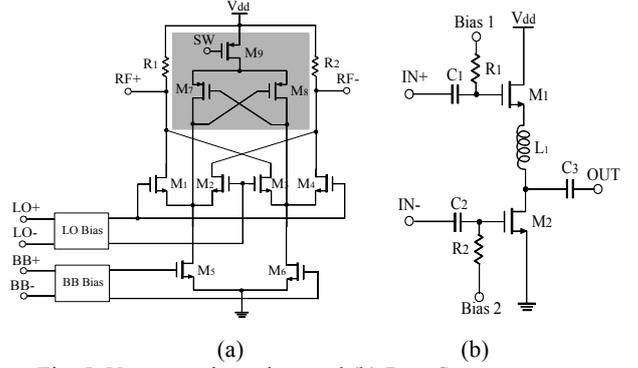

(a)  (b)
Fig. 5. Upconversion mixer and (b) D-to-S converter.

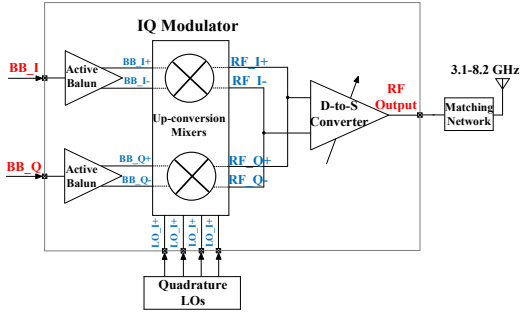

Fig. 3. Block diagram of the TRPC-UWB TX front-end.

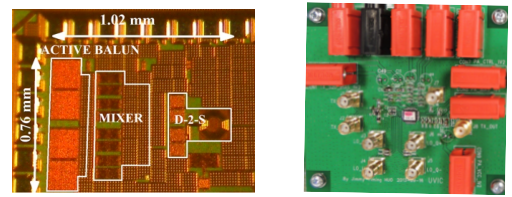

Fig. 6. Micrograph of transmitter chip and test PCB board.

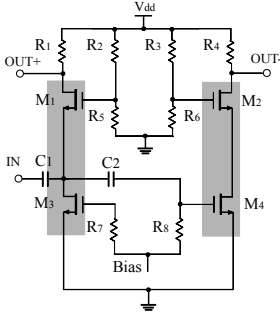

Fig. 4. Wideband balun with amplifying stages highlighted.

### 3.1. Wideband active balun

Passive balun normally occupies large physical area and introduces front end loss which increases NF, therefore a low noise active balun is designed to transform the single-ended baseband signal to differential ones. As depicted in Fig. 4, the poly resistors provide biasing, and the input signals are separated into two paths. In the first path, M1, M3 and R1 form a common-gate (CG) amplifier with nMOS current source, and it has a positive voltage gain. In the second path, a cascode structure is configured. M4 is the input device configured in the CS stage, and M2 forms the CG stage. The two paths are separated by C2 which functions as both DC blocking and phase/amplitude compensation. By using this topology, not only the noise but also the distortion of the CG transistor is cancelled [9].

### 3.2. Up-conversion mixer

Up-conversion mixer has very critical impact on the entire system EVM. In this design, the double-balanced Gilbert is designed thanks to the merits of low even-order distortion products, high input second-order intercept point (IIP2), and good isolation among ports. As described in Fig. 5(a), an up-conversion mixer consists of input trans-conductance stage, local oscillator (LO) switches, and passive loads. A quasi-differential pair with the source touched to AC ground consists of the trans-conductance stage as this topology can achieve a better the third-order intercept point (IP3). PMOS transistors M7, M8 and M9 dynamically injects the current to M5 and M6 only when the zero-crossing of the LO signal happens. Therefore the LO signal is biased at a voltage which makes only M3 conduct during the time of zero-crossing. The current injection technique significantly reduces the flicker noise translated to the RF output, and it also increases the bias current of M5, M6, without changing the bias current of M1, M2, M3, and M4 [10]. As a result, IP3 performance is also improved.

### 3.3. Differential to single-ended converter

The differential to single ended (D2S) converter finally combines the differential RF signals to the single ended ones with a tunable gain controlled by the biasing voltage. By using the current sharing technology, the maximum current consumption is reduced to no more than 4 mA. As shown in Fig. 5(b), the inductor $L_1$ functions as a peaking element resonant with the parasitic capacitor at the output of D2S.

## 4. Fabrication and measurement results

The transmitter front end chip is fabricated in a 130-nm CMOS process as shown in Fig. 6. The total area is around 0.55 mm$^2$, and the TRPC-UWB PCB board is designed in F R-4. For the verification, as shown in Fig. 7 (c), Tektronix AWG 7052 arbitrary waveform generator (AWG) is programmed to generate the baseband TRPC trains at a symbol rate varied from 10 to 300 Msps as specified in Table 1. With a 1.2 V power supply, the entire chip consumes a maximum current of 24.5 mA when operating at the highest speed mode using 7.884 GHz carrier. From the spectrum measurement, Fig. 7(b) shows the TRPC-UWB transmitter working at low speed mode under 3.827 GHz carrier frequency, while Fig. 7 (c) presents the case of high carrier frequency at 7.884 GHz when it achieves 250 Mbps data rate. They both comply with the FCC emission mask. Fig. 8(a) shows the time domain measurement when the transmitter works at 10 Mbps mode. The green waveform indicates the baseband singals while the yellow one represents the TRPC-UWB TX output. In Fig. 8(b), $N_p$ is reduced to 4, and $T_p$ is shortened to 0.85 ns, which leads to a higher speed to 100 Mbps. Furthermore in Fig. 8(c), in order to enable a higher speed at 300 Mbps, $N_p$ equals to 2, and $T_p$ is reduced to 0.85 ns so that a reasonable symbol error rate (SER) can be maintained. On the aspect of RF test, the measured carrier leakage suppression is 37.1 dBc at 3.71 GHz and the single sideband suppression (SSBS) is 28.9 dBc. Over the entire wide operating bandwidth, the carrier suppression is better than 31.6 dBc while the SSBS is higher than 22.4 dBc. Finally, the energy efficiency $E_d$ is defined using the equation below, and its unit is pico-joule per pulse

$$E_d = \frac{\text{Average Power Consumption in One Duty Cycle}}{\text{Total Number of Pulses}}. \quad (10)$$

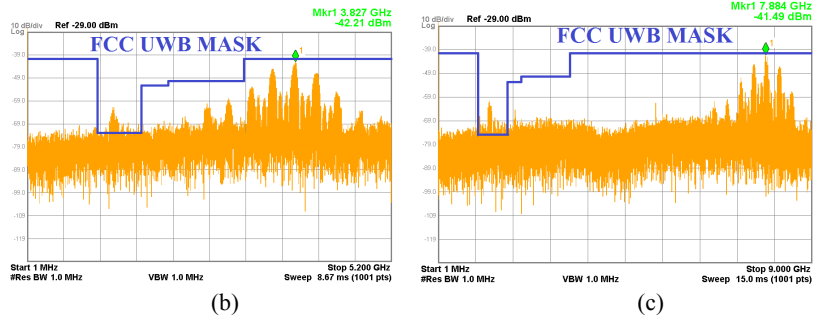

(a)                      (b)                      (c)

Fig. 7. (a) TRPC-UWB transmitter lab test set-up, (b) TRPC-UWB TX output modulated with 3.827 GHz carrier frequency, working at 10 Mbps data rate, and (c) operating with 7.884 GHz carrier frequency, working at 250 Mbps data rate.

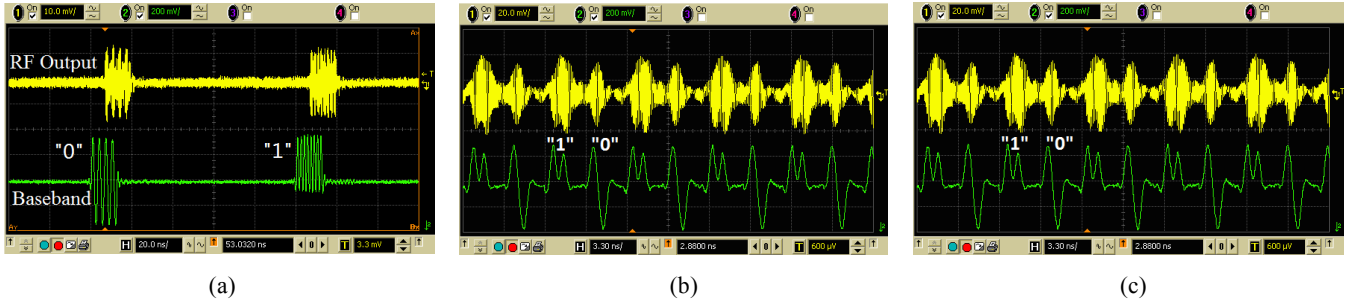

(a)                      (b)                      (c)

Fig. 8. Measured time-domain waveforms of (a) TRPC baseband signal and TRPC-UWB TX output 10 Mbps mode with 3.827 GHz carrier, (b) 100 Mbps mode with 7.88 GHz carrier frequency, and (c) 300 Mbps mode with 7.884 GHz carrier frequency.

The best energy efficiency of 38.4 pJ/pulse is achieved when TRPC-UWB transmitter works at 250 Mbps mode with $N_p$ equal to 3.

## 5. Conclusion

This paper has presented a UWB transmitter based on the novel TRPC modulation scheme. Through unveiling the relationship between data rate, pulse cluster characteristic and FCC UWB emission limit, a CMOS wideband I-Q modulator based TRPC-UWB transmitter has been designed and verified. It achieves a variable data rate from 10 to 300 Mbps over a very wide operation frequency, with good carrier leakage suppression, sideband suppression and linearity, while consuming low power and achieving a very good energy efficiency.

**Acknowledgement**

The authors would like to express thanks to CMC Microsystems, Natural Sciences and Engineering Research Council of Canada for support of this project, and Adrian Tang from JPL, NASA, for valuable discussion.